**Article Type:** [Research Article]

**Title**: [The aftermath of the Covid pandemic in the forest sector: new opportunities for emerging wood products]

**Authors:** [Mojtaba Houballah[1,7], Jean-Yves Courtonne[2], Henri Cuny[3], Antoine Colin[3], Mathieu Fortin[4], Jean-Baptiste Pichancourt[6], Francis Colin[1]]

**Institutions:** [

[1] Université de Lorraine, AgroParisTech, INRAE, Silva, 54000, Nancy, France

[2] STEEP team, Univ. Grenoble Alpes, CNRS, Inria, Grenoble INP, LJK, 38000 Grenoble, France

[3] Institut national de l'information géographique et forestière, département d'analyse des forêts et des haies bocagères, 1 rue des blanches terres, 54250, Champigneulles, France

[4] Canadian Wood Fibre Centre, Canadian Forest Service, Natural Resources Canada, Ottawa, Ontario, Canada

[6] Université Clermont Auvergne, INRAE, UR LISC, F-63172 Aubiere, France

[7] Laboratoire des Sciences du Climat et de l'Environnement, LSCE-IPSL (CEA-CNRS-UVSQ)

]

**Corresponding Author:** [Mojtaba Houballah. 14 Rue Girardet, 54000 Nancy, mojtaba.houballah@gmail.com]

**Abstract:** [

**Context**

Over the last decade, the forestry sector has undergone substantial changes, evolving from a post-2008 financial crisis landscape to incorporating policies favoring sustainable and green alternatives, especially after the 2015 Paris agreement. This evolution was drastically disrupted with the advent of the COVID-19 pandemic in 2020, causing unprecedented interruptions in supply chains, product markets, and data collection. Grasping the aftermath of the COVID-19, regional instances of the forest supply chain sector need synthetic pictures of their present state and future opportunities for emerging wood products and better regional-scale carbon balance. But given the impact of COVID-19 lock-down on data collection, the production of such synthetic pictures has become more complex, yet essential. This was the case for the regional supply chain of the Grand-Est region in France that we studied.

**Aims**

For this study, our aim was to demonstrate that an integrated methodology could provide such synthetic picture even though we sued heterogenous sources of data and different analytical objectives: i.e. (1) retrospectively evaluate the aftermath of COVID-19 pandemic on the supply chain outcomes within the forestry sector; and then (2) retrospectively explore possible options of structural change of regional supply chain that would be required to simultaneously recover from COVID-19 and transit to new objectives in line with the extraction of new bio-molecules from wood biomass, and with the reduction of the regional scale carbon footprint (in line with the IPCC Paris Agreement)

**Methods**

To achieve this, our methodological approach was decomposed into three steps. We first used a Material Flow Analysis (MFA) recently conducted on the forestry sector in the Grand Est region to establish a Sankey diagram (i.e. a schematic representation of industrial sectors and biomass flows along the supply chain) for the pre-Covid-19 period (2014-2018). Then we compared pre-Covid-19 Sankey diagram to the only source of data we could access from the post-Covid-19 period (2020-2021) in order to estimate the impact of Covid-19. Finally, we used as input the reconciled supply chain model into a consequential Wood Product Model (WPMs), called CAT (carbon accounting tool) in order to compare three prospective scenarios: (1) a scenario that projected 2020-2021 Covid-19 conditions and assumed pre-Covid-19 business as usual practices, (2) a scenario illustrating the consequence or rerouting some of the biomass to satisfy the expected increase in pulp and paper production to satisfy the needs of the industry after Covid-19, and (3) a scenario that explored new opportunities in term of extraction of novel bio-molecules by the emerging biochemical wood industry. For every scenario we also evaluated the regional carbon gains and losses that these changes implied.

**Results**

Our study conducted a detailed analysis of the impacts of the COVID-19 pandemic on the forestry sector's supply chain in the Grand Est region, using a dynamic and integrated Wood

Product Model. We found significant disruptions during the pandemic period, with notable declines in industrial wood chips and timber hardwood production by 41.8% and 40%, respectively. Conversely, there were substantial increases in fuelwood, timber sawdust, and timber softwood, rising by 14.15%, 44.23%, and 15.29% respectively. These fluctuations underscore the resilience and vulnerabilities within the regional wood supply chain. Our findings also emphasize the potential for strategic rerouting of biomass flows to meet changing industry demands, which could play a crucial role in supporting the sector's recovery and adaptation to post-pandemic conditions.

**Discussion and conclusion**

In addition, our study recognizes the limitations of the current approach combining MFA and WPM and suggests potential areas of enhancement. Ultimately, our findings shed light on the need to develop more integrated analytical methods to provide useful synthetic pictures of regional scale supply chains, when there is a need to adapt it to evolving situations and complex data landscapes.

1. INTRODUCTION

[

The global forestry sector was severely impacted by the decreased demand for construction wood following the 2008 financial crisis (Ma et al., 2009, Karsenty et al., 2010, Sayer et al., 2012, Gale et al., 2012, Canova and Hickey 2012). In response to this decline, and in line with the climate goals established in the 2015 Paris agreement, national and international institutions embarked on economic revitalization and adaptation policies. These policies aimed to increase the use of wood biomass for building a greener society, with the goal of substituting fossil fuel-based products and energies with timber and other wood biomass-extracted molecules (EU

Forest Strategy 2013, Mandova et al., 2018, IEA 2019). These strategies fostered improvements in supply chains and encouraged the recycling of wood by-products to produce energy (Braghiroli and Passarini, 2020) and extract promising molecules from various species (Ben-Iwo et al., 2016, Pichancourt et al., 2021). For example, since 2012, the European policymakers have been promoting the transition from first-generation biofuels to advanced biofuels. These advanced biofuels are produced from residues and wastes of forest-based industries, and their production has been incentivized (Council of the European Union, 2018, Jonsson et al., 2021).

In the decade following the 2008 financial crisis, the global demand for other wood-based products also increased (A Blueprint for the EU Forest-Based Industries 2013, Wolfslehner et al., 2016, Kardung and Wesseler 2019). Given the competition for forest resources, the structure of regional supply chains was under stress and had to adapt to this increasing diversity in the demand (Rozhkov et al., 2020, Kano et al., 2022). This situation outlined the need for assessment tools to identify adaptation pathways that comply with the new green-house gas emissions targets set by the 2015 Paris agreement (Pichancourt et al., 2018).

In response to the Covid-19 outbreak in the 2020-2021, many governments decreed a series of lockdowns which significantly affected the operations along the supply chains, vendors, and product commercialization (Bulin and Tenie 2020, Bhandari et al., 2021, Laudari et al., 2021, Chirwa et al., 2021). Initially, lockdowns caused a break in supply chains and delayed the final uses of wood products. Globally, timber demand declined in 2020 and 2021 (Stanturf and Mansuy 2021). However, after the lockdown period, the demand for wood products surged, overburdening supply chains due to the high demand from industries such as construction and manufacturing (Buongiorno 2021), and from paper and paperboard packaging due to an increase in online orders (Liu et al., 2020). In the Grand-Est region of France, the pulp industry has shifted its production from newsprint to recycled paperboard in response to the growth of e-commerce (Norske Skog 2020, 2021a, 2021b). As of 2022, the disruptions to the functioning of the forest sector are still evident and likely to continue even if COVID-19 is no longer considered as a global health emergency (https://news.un.org/en/story/2023/05/1136367).

Policymakers in various countries have taken action to mitigate the impact of COVID-19 on wood supply chains (Desson et al., 2020, Wahyuni and Wiati 2021). However, as suggested by the Food and Agriculture Organization (FAO 2020), additional research and methods need to be developed to guide decisions aimed at a green recovery of the economy based on the forest sector. A consensus is emerging around the idea that without a unified understanding of the significance of our forestry supply chain for the various dimensions of our societies, we cannot fully evaluate the effectiveness of forest sector policies and the impacts of economic disturbances, such as Covid-19, on the supply chain (Piquer-Rodríguez et al., 2023). While significant progress has been made in representing and understanding forestry supply chains across these various dimensions, assessment tools are still not fully adapted to the current situation. COVID-19 presents an opportunity to meet this challenge.

This paper introduces a novel methodology that integrates a dynamic wood product model with material flow analysis, designed to offer a comprehensive and adaptable framework for analyzing forestry supply chains. By employing data from 2014-2018 as a stable reference period, we avoid the transient complexities introduced by the pandemic, thus ensuring clarity and consistency in our analysis. We apply this methodology to evaluate three scenarios within the Grand-Est region: a baseline scenario representing the state of the supply chain before the pandemic, a COVID-19 scenario to assess the immediate impacts of the pandemic, and a prospective scenario that explores the potential of the emerging biochemical wood industry. This approach not only assesses the direct effects of the pandemic but also examines possible structural adaptations that could enhance sustainability and economic resilience in the long term. Through this study, we aim to demonstrate the utility of our integrated approach in providing dynamic, responsive insights into forestry supply chains, capable of informing policy and operational decisions in the face of both existing and unforeseen challenges. The findings underscore the importance of developing flexible, robust assessment tools that can help navigate the ongoing and future transformations in the forestry sector.]

## 2. METHODOLOGY

[

### 2.1 Wood product models

Wood product models (WPMs) allow for schematic representations of the forest sector and they are widely used in carbon accounting. These models can be used for impact assessments that simultaneously incorporate biomass, economic, and environmental criteria, in relation to various external shocks or longer-term adaptation policy scenarios (Brunet-Navarro et al., 2016). Because wood biomass units can be converted into $CO_2$ equivalent units (COP17, 2011), existing WPMs can provide joint assessments of wood biomass and carbon accountability across an entire supply chain for a specific regional forest sector (e.g., Brunet-Navarro et al., 2016, Pichancourt et al., 2018, Pichancourt et al., 2021). Some of these WPMs can further analyze changes in relation with harvested species, the number of different biomass components and available by-products, and changes in the locations and time windows of extraction or transformation (Brunet-Navarro et al., 2016). A schematic representation of supply chains, even at the regional level, requires a great deal of information about who the stakeholders are and the wood flows between them.

The challenge, however, is that this information comes from heterogeneous data sources. This heterogeneity often gives rise to statistical problems when trying to estimate supply-chain level biomass production metrics (Innovating for Sustainable Growth: A Bio-economy for Europe, 2012; Grassi et al., 2017; Scarlat et al., 2015). For example, errors can occur when the sources of data cannot be specified in advance (Brunet-Navarro et al., 2016). Some data sources, like information on forestry by-products from different species, data from post-consumer wood products, and urban greenery management data (Camia et al., 2018; Sileshi, 2014), or data linking the diversity of biomass products to the technical limits used for their mobilization (Hamelin et al., 2019), are particularly prone to these problems.

## 2.2 Material flow analysis

Material Flow Analysis (MFA; Fischer-Kowalski, 1998) has proven a systematic method for reconciling heterogeneous data sources when it comes to modeling biomass flows along supply chains, which can be represented as a Sankey diagram (Courtonne et al., 2019). With integrated data reconciliation (see section 2.3.2), Material Flow Analysis (MFA) models, when applied to forestry supply chains, have demonstrated significant potential. They effectively illustrate the structural rules that govern the stocks and flows of wood biomass across various spatial and temporal dimensions, both within and beyond a specific area (Schaffartzik et al., 2014; Lupton and Allwood 2017, Marques et al., 2020; IRENA 2019 a, b; AF Filières project 2015; Avitabile and Camia, 2018; Camia et al., 2018). These models have been utilized to depict supply chains at varying scales, including both national and regional scales (Binder et al., 2004; XERA, 2017; Koen et al., 2019; Parobek et al., 2014). Additional studies addressing the increased demand for forest biomass and its impacts can be found in (van den Born et al., 2014; Börjesson et al., 2017). Essentially, MFA and WPM are complementary tools. MFA can generate standardized Sankey diagrams of supply chains (Courtonne et al., 2019) which provide the information to create a schematic representation of the supply chain in a WPM. MFA is not meant for dynamic updates because they use data from multiple sources. They typically focus on a single flow of metric (e.g., wood biomass) and currently cannot infer other crucial metrics, such as carbon equivalent content from wood. On the other hand, WPMs can explore the changes induced by the supply chain structure on biomass flow dynamics and carbon balance, but they cannot reconcile heterogeneous data.

## 2.3 An integrated approach

The methodology used in this paper is based on the coupling of an existing WPM called CAT (Carbon Accounting Tool, see Pichancourt et al., 2018) with a French MFA called AF-Filières (Courtonne et al., 2015, 2019). This methodology can be decomposed into three steps (Figure 1).

### 2.3.1 Step 1: Data collection and formatting

In France, data on the forest sector are generated by various public organizations. The National Institute of Geographic and Forest Information (IGN) publishes forest data (forest area, standing volumes, annual growth, mortality and felling…) through its National Forest Inventory (NFI), while the regional interprofessional association for the forest-based sector (FIBOIS) and the regional management institute for food, agriculture, and forests (DRAAF) independently provide economic network data (see Table 1). These entities use different sampling protocols and scales of observation, necessitating data reconciliation to reconstruct the biomass flow in the regional forest-based sector's biomass flow model.

A significant challenge arises from the varying units for quantifying wood biomass across its transformation process among these sources. For consistency, we converted all the units in cubic meters ($m^3$) across the entire supply chain (Weimar, 2011). The cubic meter unit also called the wood fibre equivalent is considered a European standard for comparative studies (Mantau, 2012; Bosch, 2015; Weimar, 2011) and it represents the volume of wood fiber in a product at the fiber saturation point, remaining unaffected by water content. It is ideal for estimate such as green roundwood. Using this unit required the calculation of a specific conversion coefficient for each wood-based product (Lenglet et al., 2017).

### 2.3.2 Step 2: Data reconciliation through MFA

In 2021, a working group leads by the DRAAF and involving several other institutes (INRAE, IGN, FIBOIS and INRIA) was launched to test the AF Filières MFA tool on the forest sector of the Grand Est region (citer le rapport de JL Matte). For that, data on different aspects of the forest sector and originating from different sources (notably NFI data of IGN for the forest, FIBOIS and DRAAF data for the economic network) were compiled and reconciled. Finally, this work has generated a representation of the Grand Est forest sector for the 2014-2018 period, with the material flows from the forest resource to and between the users of this resource. The Sankey diagram thus reconstructed can be visualized here: https://terriflux.com/portfolios/ForetBoisGrandEst/index.html.

### 2.3.3 Step 3: Using the MFA output in a WPM

We incorporated the Sankey diagram of the Grand-Est supply chain into the Wood Product Model (WPM) CAT (Fortin et al., 2012; Fortin et al., 2014; Pichancourt et al., 2018). Based on forest resource data and a scheme of wood biomass allocation between the diverse stakeholders this tool can calculate the stocks and flows of carbon along the supply chain.

For this study, the forest resource data used in CAT were those calculated for the Grand Est region from the NFI for the 2014-2018 period, while the allocation scheme of wood biomass allocation between the diverse stakeholders was the one derived from the Sankey diagram descried above.

CAT distinguishes various carbon pools—aboveground and belowground living biomass, dead organic matter, and harvested wood products (HWP), both in use and deposited at solid waste disposal sites (SWDS). Importantly, CAT allows us to calculate the supply chain's carbon balance following the IPCC guidelines for national greenhouse gas inventories from Land Use, Land-Use Change and Forestry (LULUCF) and waste sectors (IPCC 2006a, b, 2014). However, its most significant contribution to our study is its ability to facilitate scenario building. CAT's flexibility enables us to modify the wood supply chain parameters, thereby allowing us to construct and analyze various scenarios reflecting potential changes in the forestry sector. This feature is instrumental in our exploration of the impacts of different policy decisions or industry trends on the regional carbon footprint and overall sustainability of the region.

## 2.2.4 Step 4: Policy Scenario Analysis and Construction for WPM

WPMs are beneficial for prospective analyses exploring future adaptation pathways, based on the current system state. This is particularly relevant to us, as we need to suggest possible COVID-19 recovery pathways but lack data covering the pandemic period. Consequently, constructing scenarios before and after COVID-19 is essential. Using the dynamic flow manager of CAT, we developed several scenarios that reflect potential shifts in biomass production or consumption. These scenarios were designed to project their impacts on flow balance and stock. The first scenario explores the impact of changes in the recycling practices within the wood panel and pulp industries. The second scenario investigates the potential for enhancing the value of bark through the introduction of a chemical actor in the regional timber sector. These alternative scenarios, along with the baseline scenario representing the pre-COVID situation, allow us to explore a range of possible futures and understand the potential impacts of different policy decisions or industry trends. The relevance of these alternative scenarios lies in their ability to provide insights into the resilience and adaptability of the forestry sector under different conditions. Our approach for factoring in the anticipated impact of COVID-19 comprised three sub-steps:

Sub-step (4.1): Establish a baseline scenario representing the pre-COVID-19 period.

Sub-step (4.2): Gather detailed data and estimators sufficient for evaluating an evolution scenario.

Sub-step (4.3): Compare results across scenarios.

In this case study, the baseline scenario was the regional forest-based sector model, created using AF Filières and integrated into CAT. This model assumed no changes in the interactions between stakeholders in the forest sector of the Grand-Est region. The precision of the data made it possible to assess two scenarios in the timber sector, where softwood and hardwood flows are mixed.

As mentioned before, the first scenario simulates a situation where wood panel or pulp industries cease using by-products from the timber sector. This resembles a recycling policy where the crushing industry only uses old panel and paper products for manufacturing new ones. This scenario translates to a significant shift in by-products' final destination. During the COVID-19 period, the forest-based sector experienced worldwide impacts. Reduced industrial production was followed by a surge in wood product demand due to economic recovery (Buongiorno, 2021). The pandemic increased demand for paper and paperboard packaging due to a rise in online orders (Liu et al., 2020). In Grand-Est, the pulp industry shifted its production system from newsprint to recycled paperboard, thereby ceasing future roundwood and by-products use. The wood panel industry, too, increased its use of recycled wood boards. However, due to data limitations from the AF Filières project, the scenario simulated the impact of a global industrial tool reconversion. The second scenario centers on bark enhancement. It considers the addition of a chemical actor to the previous COVID-19 scenario to valorize bark produced in the regional timber sector (Feng et al., 2013). This scenario calculates the maximum wood extractives deposit that can be allocated to another wood valorization alternative. The rationale is to simulate the development of a chemical wood industry as part of greener forest exploitation (Maarit et al., 2018). Such an approach has already begun at the regional level in Grand-Est, with new industries emerging under the chemical wood category.

]

1. RESULTS

[

**Retrospective impact of COVID19 on supply chain outcomes**

A comparison between pre-COVID-19 data (2014-2018 MFA reconciliation data) and COVID-19 data (2020-2021 FIBOIS observatory) reveals significant changes in production and supply chain outcomes resulting from the COVID-19 lockdown (Table 2). Certain classes of wood products experienced substantial decreases, such as forest wood chips for energy (-14.2%), timber industrial wood chips (-41.8%), and timber hardwood (-40%). Conversely, other products saw significant increases, including fuelwood (+14.15%), timber sawdust (+44.23%), and timber softwood (+15.29%). The production of bark byproducts and sawdust from the timber industry remained relatively unaffected. Whereas, the same outcomes could not be accessed for the crushing wood sector, as this strategic information involves only one company in the region.

**Effect of the evolving recycling policy and COVID19 related increase in paper production on the consumption of by-products by the crushing industry**

We could not get regional-scale COVID19 data for the crushing industry (Table 2). However, based on public data and conservative assumptions based on Liu et al., (2020) we could predict what may have been the production of the increased demand in pulp and paper due to COVID19. All other things considered, CAT WPM projections (Figure 3) predict *ceteris paribus* that approximately 69,000 m$^3$ of wood chip and 27,000 m$^3$ of sawdust (i.e., 96,000 m$^3$ or 11% of their biomass) may be re-routed per year, in order to support the changing needs in the crushing industry to produce more paper during and after COVID19. Bark may not be rerouted, and still be burned for the energy sector. This rerouting from the energy sector to the crushing industry represents potentially 24,000 tons of carbon that has been saved from burning by the energy sector and released in the atmosphere. Given the rerouting is likely temporary, quantities of wood chip and sawdust could be re-allocated to other industrial sectors. Similarly, if needed. furthermore, 19 % of the regional timber is expected to be exported before any transformation by the regional industries (i.e., 131,000 tons of carbon), and an anecdotal quantity of by-products is exported from the region (as industrial wood chips, sawdust, and bark), corresponding to an extra 8,000 tons of carbon biomass leaving the region.

**Effect of introducing extensive chemical wood industries**

Given these constraints, we predict in a second scenario, that 77,000 m$^3$ (this number being hypothetically defined) of wood biomass that could be rerouted from the energy sector to extract bio-molecules for the new emerging biochemical industry (Figure 4). This projection is based on a conservative estimate of a maximum of 39 000 tons of extractive deposit that can be processed from the main species harvested in the region (i.e., sessile oak, pedunculate oak, European beech, silver fir, Norway spruce, and Douglas fir, https://agriculture.gouv.fr/infographie-la-filiere-foret-bois-en-france). For the carbon balance, this rerouting corresponds to a maximum of 20,000 tons per year equivalent in carbon that is saved from burning and released in the atmosphere.

If both scenarios presented here can coincide, the lack of wood biomass could be resolved theoretically by adding 96 000 m$^3$ of by-products (wood chips and sawdust), as the needs by the crushing industry may be temporary. In both cases, the deposit predicted by CAT software represents the industrial limit and an indicator of the emergence of a chemical wood sector.

]

## 4. DISCUSSION

[

Our research delved into the substantial effects of COVID-19 on the wood supply chain, revealing distinct shifts in the production of various wood products. The wood crushing industry demonstrated remarkable adaptability, meeting an increased paper demand during the pandemic with the potential for significant carbon savings via strategic rerouting of by-product. Additionally, we investigated the feasibility of diverting wood biomass from the energy sector to support the emerging biochemical industry. The uptick in demand from the crushing industry might be balanced by increasing the utilization of by-product, indicating a key evolution in the wood industry in response to shifting needs and policies.

In this context, the Carbon Accounting Tool (CAT) emerges as a significant component in evaluating the impact of COVID-19 on regional supply chains. CAT's comprehensive approach to carbon accountability in managed forests can be integrated into the Material Flow Analysis (MFA) and Wood Product Model (WPM) tools. This integration is particularly relevant when reconciling heterogeneous data sources, as in the case of the French Grand-Est region's wood supply chain. The study by Wang et al. (2022) underscores the potential of tools like CAT in enhancing these models. Furthermore, the shift towards digital and low-carbon economies, highlighted by Lee (2021) and Wang, Wu, and Chen (2021), underscores the importance of trust and credibility in global supply chains. The practical perspective provided by Hernández (2019) on analyzing forest wood supply chains for round-wood production considering technical constraints could potentially be integrated with CAT to provide a more comprehensive view of carbon accountability. In essence, CAT's comprehensive and adaptable approach to carbon accountability makes it a crucial tool in the pursuit of sustainable wood supply chains, especially in the context of evaluating the impact of COVID-19 on regional supply chains.

**Impact of Recycling Policies on the wood industry Carbon Footprint**

In this article, we're examining the Material Flow Analysis (MFA) to determine how modifications in production or processing can influence the carbon footprint. In France's Grand-Est region, timber industry byproducts such as wood chips and sawdust are repurposed in the creation of wood panels. However, these by-products could alternatively be utilized in the energy sector. If the crushing industry, responsible for producing wood panels, decided to shift from using by-products to recycling old wood panels, it's likely these waste materials would be redirected towards energy production. This shift, while potentially beneficial for the environment, could also have significant economic implications. The costs associated with changing production processes, potential savings from using recycled materials, and the economic value of the energy produced from waste materials would all need to be considered.

The role of government policy and regulation in nudging such a shift cannot be overstated. From incentives for recycling to regulations on waste disposal and policies that champion renewable energy, all these could sway the decisions of the crushing industry. Yet, we must also acknowledge potential technological hurdles. The viability of recycling old wood panels or using waste materials for energy production could hinge on the availability and evolution of suitable technologies.

A study by Kim and Song (2014) underscored the lifespan of the produced materials, illustrating that energy products from combustion are consumed within a year, while wood panels boast a lifespan of 14 years and can undergo multiple recycling processes. This recycling journey could stretch a single panel's service life to a whopping 224 years and sequester 409 kg of $CO_2$-equivalent carbon. Consequently, shifts in recycling policies could potentially trim the carbon footprint (a point also supported by Ng et al., 2014) of the crushing industry by supplying more waste materials for energy production and prolonging carbon sequestration in recycled panels.

While the situation in France's Grand-Est region offers a compelling case study, it's crucial to remember that similar dilemmas between recycling and energy production are being grappled with by industries and regions globally. By comparing the strategies adopted in different contexts, we can glean deeper insights into the most effective tactics for curbing carbon footprints and fostering sustainable practices.

**The Potential of Forest Biomass in the Transition to a Greener Economy**

The International Energy Agency (IEA) has proposed pathways to keep global warming below 2°C, and a key strategy is to increase the use of renewable energies like biomass products. Forest biomass, in particular, is a versatile carbon source that could replace fossil fuels, offering potential for various sectors, especially in the bio-economy where it could be transformed into a range of molecular products. This underlines the appeal for a transition to an eco-friendlier chemical economy, where products currently derived from petrochemicals could be replaced with renewable alternatives, or even newly designed ones. This transition involves developing industries that harness forest and wood extracts, but the impact on the global wood supply chain needs careful assessment. As a case study, we explore the potential consequences of introducing a chemical wood industry in the Grand-Est region of France. We highlight how this new industry could affect other sectors like energy and crushing, providing valuable insight to inform decision-making processes for regional management, which presents both challenges and opportunities for management. The requirement of rerouting a projected 77,000 m3 of wood biomass from the energy sector to this newly emerging industry necessitates careful planning and coordination to ensure the needs of all sectors are met, while minimizing disruption. This shift also impacts carbon emissions, potentially saving up to 20,000 tons per year of carbon that would otherwise be released by the energy sector, thus contributing to sustainability efforts (P. Dees et al., 2023). However, this requires diligent tracking and reporting of carbon emissions. Moreover, managing the emergence of the chemical wood industry involves navigating new regulatory (Sprecher and Kleijn 2021) and market landscapes, fostering innovation, and building key stakeholder relationships. Resource management is also crucial, with the proposed use of 96,000 m3 of by-products potentially addressing any deficit in wood biomass, emphasizing the importance of effective by-product management and recycling strategies (Dalalah et al., 2022). In addition, the potential for the reallocation of resources, such as wood chips and sawdust, from the crushing industry to other sectors, including the chemical wood industry, calls for robust cross-sector coordination and agile response to shifting demands. Hence, the management of this transition demands a forward-thinking and adaptable approach, centered on sustainability and cross-sector cooperation.

**Adapting Wood Supply Chain Management in Response to the COVID-19 Scenario: Challenges and Strategies**

The COVID-19 pandemic has initiated substantial changes in the wood supply chain, requiring a reassessment of management tactics to enhance resilience and sustainability in the industry ([link1](link1), [link2](link2), FCBA 2020 in French). Ocicka et al. (2022) assert that these effects have not been evenly distributed across all wood products. This aligns with observations from the US, where the production of certain items, including forest wood chips for energy and industrial timber chips, has seen a downturn. This necessitates a revamping of sourcing methods, production schedules, and inventory control to counteract supply interruptions ([Link 3](#)). On the other hand, a rise in the production of other products, such as fuelwood, timber sawdust, and softwood lumber, has been noted. While this increase offers opportunities for expansion and diversification, it also brings challenges associated with managing the surge in demand, including price regulation and sustainable procurement ([Link 4, Link 5,](#) Businesscoot, 2021).

These changes are in line with those reported in scholarly sources, emphasizing the industry's dynamic response to external disruptions and the importance of adaptive management strategies. The pandemic has also driven an increase in paper production, necessitating a redirection of wood chip and sawdust resources from the energy sector to the pulp industry. This reallocation presents opportunities for both sectors, but it also calls for an evaluation of the long-term sustainability of this shift and the formulation of alternative sourcing or usage strategies.

The findings reported in the emerging research echo these changes, underscoring the need to evaluate the long-term impacts of these shifts on the wood supply chain. Moreover, the rise of extensive chemical wood industries, resulting from the diversion of wood biomass from the energy sector, poses new challenges that demand careful planning, sustainable sourcing of raw materials, and management of potential competition between sectors for the same resources. This issue could potentially be alleviated by incorporating by-products, thus underscoring the role of waste management and recycling strategies in the wood supply chain. This viewpoint is consistent with current research, which stresses the necessity for innovative and adaptable management strategies that can respond to evolving situations and seize emerging opportunities.

**Methodological limitations, room for improvements, and conclusions**

The current data used for generating the Sankey diagram, a type of flow diagram, only allows us to categorize wood into general groups, such as softwood and hardwood. It does not provide specific details about individual species of wood. This is a limitation because different species, like oak and beech, have unique characteristics and uses within the Grand-Est region. To improve this, we need to refine our methodology and use more detailed datasets. Moreover, the Sankey diagram, as constructed by AF Filières, provides a complex view of all wood flows within the Grand-Est region, including imported and locally used wood. However, it does not clarify how imported wood or products affect local uses. The flow values in the diagram represent a mix of locally produced and imported wood, which may not accurately depict the actual industrial production in the region (Lambin and Meyfroidt 2011). This blending of local and imported wood makes it challenging to distinguish between the two (Tenenbaum 2008). We need to address this in our future approach. To enhance our understanding of wood flows, we should work towards identifying what portion of each product comes from local industries versus imports. Similarly, it would be beneficial to determine what percentage of exported products are locally produced versus imported (Nölke and Vliegenthart 2009). This detailed understanding of wood flows can significantly aid forest certification efforts. Knowing the exact proportions of local versus imported wood in a product can help in accurately labeling products based on their origin and sustainability. This will support the certification process, which relies on understanding the wood supply chain.

While the methodology applied in this study offers valuable insights, there is potential for refinement, particularly concerning the estimation of error propagation. Multiple sources of uncertainty exist within this approach, stemming from the underlying data used, the equations employed in the AF Filières project as per Courtonne (2017), and the simulations initiated by the CAT tool, as referenced in Pichancourt et al., (2018, 2021). Thus, it becomes crucial to scrutinize these uncertainties, gauge their potential impacts, and understand how they may ripple through the model. This process of error analysis and quantification is essential for ensuring the reliability and accuracy of our findings.

With the method we present, we hoped to take the first step towards a larger goal: making Material Flow Analysis (MFA), wood product models (MFA), and industrial and emissions research in France more accessible and well-informed. Understanding and quantifying the flow of wood-based products (including the one associated with the emerging bio-molecular industry) and their associated carbon footprints are important for understanding the impact of COVID19 on the forestry industry. This is important because countries need to evaluate the best pathways to recover from COVID19, from an economic standpoint but also to meet national commitments to the Paris agreement. The first step involves understanding material flow and estimating its impact on current emission levels. In this setting, having standardized tools and methodologies is critical. It allows for carbon footprint comparisons, statistical analyses, productive discussions, coordinated reduction strategies, and consistent reporting of mitigation actions' results. The tools used in our methodology are available online, making them accessible to research labs worldwide. This accessibility encourages comparative research between different labs. By introducing this methodology, following Brunet-Navarro et al., (2016), we are broadening research opportunities and pushing the conversation about the best methods to estimate the impact significant systemic shocks (such as COVID19) on multi-functional forestry material flow and carbon balance footprint.

]

## ACKNOWLEDGMENTS

[

This research has been funded under the ExtraFor_Est project financed from five entities (MAAF, FEDER Lorraine, LABEX Arbre, ADEME, and the region of Grand-Est). Moreover, we gratefully acknowledge support from DRAAF, IGN, Fibois, INRIA, and INRAE for their contribution to the data collection used in this article. Data reconciliation was facilitated by the collaboration between various French state organization (IGN, FIBOIS, DRAAF) and research institutes (INRIA and INRAE)

]

# SUPPORTING INFORMATION

**Supporting Information**

Supporting information is linked to this article on the *JIE* website:

**Supporting Information S1:** This supporting information provides…

**Supporting Information S2:** This supporting information provides…

**Figure Legends**

[

*Figure 1. The four steps of the methodology used to evaluate the impact of COVID-19 on the regional forestry supply chain: **Step 1:** data collection and formatting; **Step 2:** data reconciliation to produce a standardized Sankey diagrams using MFA model; **Step 3:** data filtering and incorporation into a WPM model; and **Step 4:** analyze the COVID-19 impact using WPM software (e.g., CAT)*

*Figure 2. The perfect blend hypothesis made explicit (from Courtonne, 2016).*

*Table 2. Comparison of the volume estimated by AF Filières and the model onto CAT with the FIBOIS observatory (all species combined). The percentages translate the gap with the FIBOIS data (reference). The values are in cubic meters of wood fiber equivalent (x 1 000). The label N/A designates not available data.*

*Figure 3. Timber sector in the Grand Est region of France (baseline scenario) and localization of the wood waste flows impacted by a change in the crushing industry (first scenario, red flows). The blue flows are the exports of wood or products from the Grand Est region. The grey flows translate the wood or products valorized in the region.*

*Figure 4. Chemical wood sector view for the timber bark in Grand Est region of France (second scenario, green flows). The blue flows are the exports of wood or products from the Grand Est region. The grey flows translate the wood or products valorization in the region.*

]

*Table 1. Overview of the data sources, methodologies, and objectives for each period of analysis, ranging from pre-COVID-19 benchmarking to post-COVID-19 prospective analysis in the regional wood supply chain.*

| Period | Data sources | Data treatment & analysis | Objective |
|---|---|---|---|
| **2014-2018** | *Multiple: DRAAF, IGN* | *Reconciliation using MFA model called AF Filière* | *Serves as a pre-COVID-19 benchmark, facilitating the analysis of COVID-19's impact and providing a foundation for prospective scenario planning.* |
| **2020-2021** | *Single: FIBOIS* | *Note reconciled & compared to pre-Covid-19 baseline scenario* | *Estimate the impact of Covid-19 on the regional wood supply chain* |
| **> 2021** | *No Data* | *Prospective analysis using WPF & CAT* | *Exploring the post-Covid-19 opportunities of the regional supply chain* |

*Table 2. Estimation of the impact of COVID19 on the different wood industrial sectors. The percentages translate the difference between the FIBOIS observatory data sampled during the COVID19 period (2020-2021) and the baseline scenario from the pre-COVID19 period (2014-2018). The values are in cubic meters of wood fiber equivalent (x 1 000). The label N/A designates no data is available.*

| Product | Pre-Covid19 data from 2014-2018 (using MFA reconciliation) | Covid19 data (2020-2021: using FIBOIS observatory) | Retrospective impact of Covid19 |
|---|---|---|---|
| **Energy sector** | | | |
| *Fuel wood* | 5 073 | 5 909 | +14.15 % |
| *Forest wood chips* | 1 496 | 1 310 | -14.20 % |
| **Crushing wood sector** | | | |
| *Wood panel* | 2 307 | N/A | N/A |
| *Pulp* | 1 469 | N/A | N/A |
| *Bark* | 215 | N/A | N/A |
| **Timber sector** | | | |
| *Industrial wood chips* | 943 | 665 | -41.80 % |
| *Sawdust* | 377 | 676 (sawdust and wood shaving) | +44.23 % |
| *Bark* | 401 | 397 | -1.01 % |
| *Sawing* | 1 487 | 1 477 | -0.68 % |
| *Softwood* | 892 | 1053 | +15.29 % |
| *Hardwood* | 595 | 424 | -40.33 % |

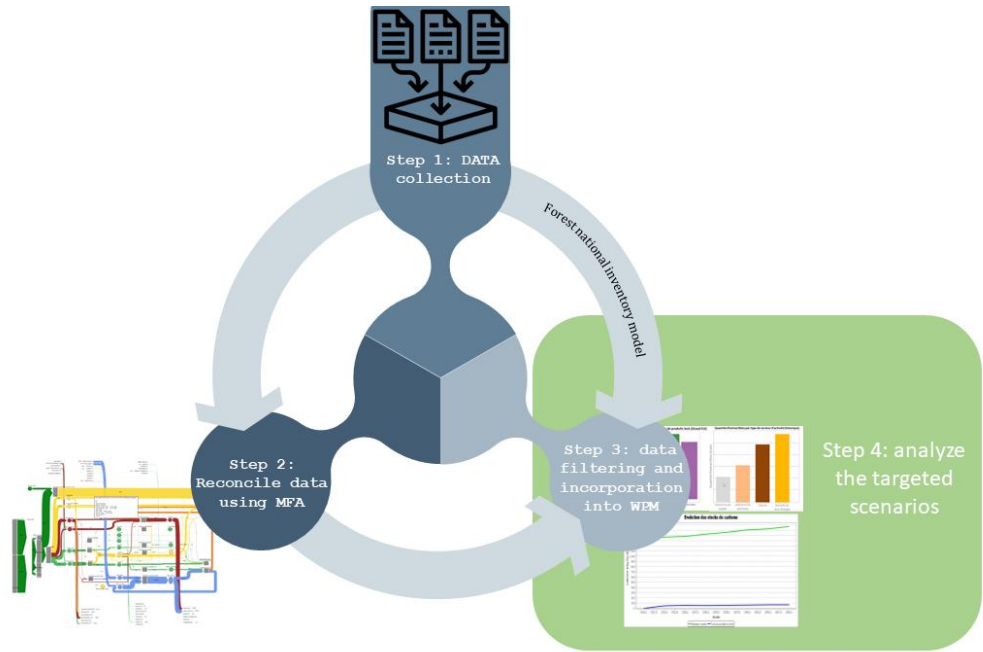

*Figure 1. The four steps of the methodology used to evaluate the impact of COVID-19 on the regional forestry supply chain:* **Step 1:** *data collection and formatting;* **Step 2:** *data reconciliation to produce a standardized Sankey diagrams using MFA model;* **Step 3:** *data filtering and incorporation into a WPM; and* **Step 4:** *analyze the COVID-19 impact using WPM software (e.g., CAT)*

**What happens in the real world:**
Production (p) is either directed to local consumption (pc) or exports (pe).
Imports (i) are either directed to local consumption (ic) or exports (ie).
Goods consumed (c) are either produced locally (pc) or imported (ic).
Goods exported (e) are either produced locally (pe) or imported (ie).

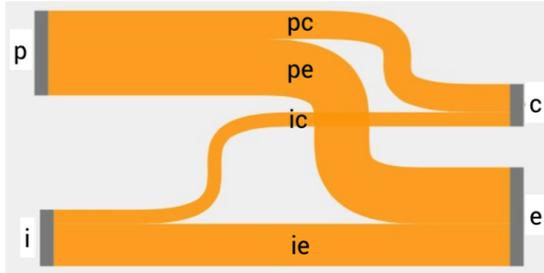

**Available data:**
Production (p), Imports (i), Supply (s), Consumption (c), Exports (e)

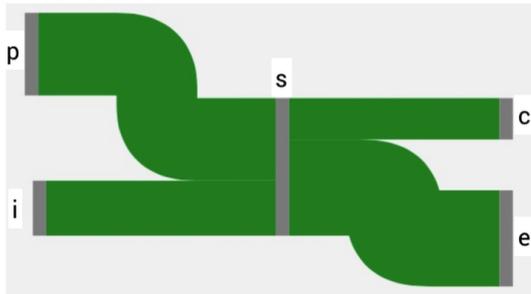

**In order to compute flows pc, pe, ic and ie, we use the perfect blend hypothesis:**

$$pc = p/s \cdot c = c/s \cdot p$$
$$pe = p/s \cdot e = e/s \cdot p$$
$$ic = i/s \cdot c = c/s \cdot i$$
$$ie = i/s \cdot e = e/s \cdot i$$

*Figure 2. The perfect blend hypothesis made explicit (from Courtonne, 2016).*

*Figure 3. Sankey diagram of the predicted wood volumes (m³) and associated carbon storage (tons of carbon) transiting every year along the regional supply chain in the French Grand Est, The diagram presents wood biomass flows based on pre-COVID19 reconciled data, supplemented by a scenario of rerouting of by-product flows to the crushing industry, to reflect the expected increase in pulp and paper during and after COVID19. Data for the crushing industry comes from DRAAF. Wood exports have two origins: directly after harvest and after first transformation.*

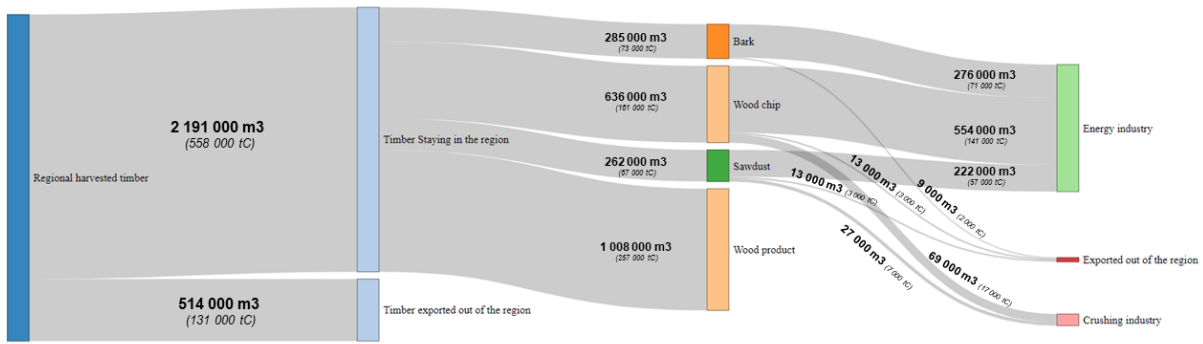

*Figure 4. a) Sankey diagram of the wood volumes transiting along the regional supply chain in the French Grand Est before COVID19 in 2014-2018. (b) CAT projections with emphasis (green flows) on the potential bark volumes that could be used for the emerging chemical wood sector. The blue flows are the exports of wood or products from the Grand Est region. The grey flows represent the flow of wood or products along the supply chain.*

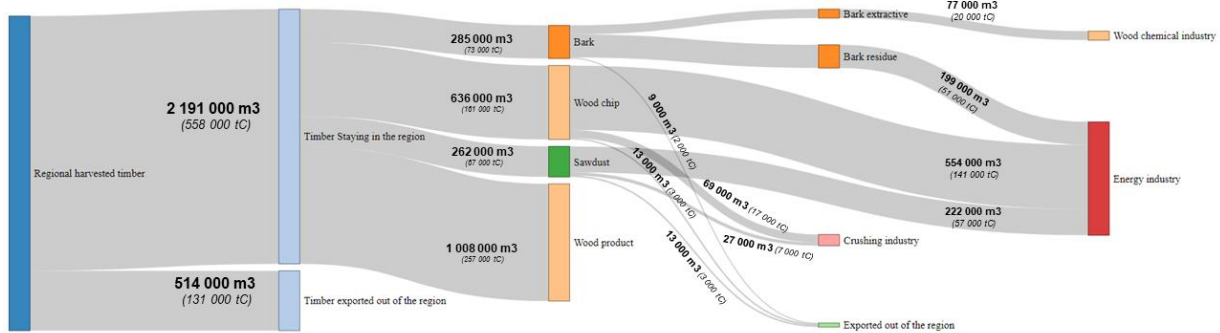

37